# DECOMPOSITIONS FOR SUM-OF-POWERS STATISTICS AND THE SAMPLE CENTRAL MOMENTS

BEN O'NEILL[*], *Australian National University*[**]

WRITTEN 15 JANUARY 2021

**Abstract**

We give some useful decompositions of sum-of-powers statistics, leading to decompositions for the sample mean, sample variance, sample skewness and sample kurtosis. We solve two related problems: computing these sample moments for a pooled sample composed of subgroups with known moments; and computing these sample moments for a subgroup using known moments from other subgroups and the overall pooled sample. Each task is accomplished via decompositions of the sums-of-squares, sums-of-cubes and sums-of-quads from which the sample central moments (up to fourth order) are formed. We give decomposition results and we implement these in a user-friendly **R** function.

DESCRIPTIVE STATISTICS; DECOMPOSITIONS; SAMPLE MOMENTS; SAMPLE MEAN; SAMPLE VARIANCE; SAMPLE SKEWNESS; SAMPLE KURTOSIS.

A problem sometimes encountered in applied data analysis is when an analyst has information on the descriptive statistics of a set of subgroups, and is required to aggregate these subgroups into a pooled data set without access to the underlying data. In this case, the analyst wishes to generate descriptive statistics for the pooled data, using only the descriptive statistics of the subgroups. A variation on this problem occurs when the analyst has descriptive statistics for the pooled sample and all but one of the subgroups that make up that pooled sample, and they are required to find the descriptive statistics of the missing subgroup. In all these cases it is necessary for the analyst to use rules pertaining to decompositions of descriptive statistics in order to obtain the descriptive statistics for the sample of interest.

For some descriptive statistics it is impossible to obtain the missing values without access to the underlying data. For example, the quantiles of the pooled sample (other than the pooled minimum and maximum) cannot generally be obtained from quantiles of the subgroups, and likewise, quantiles of a missing subgroup cannot be obtained from the quantiles of the pooled sample and the other subgroups. The sample minimum and maximum can easily be obtained for the pooled sample from the subgroups, but the sample minimum and maximum of a missing subgroup cannot be obtained from the pooled sample and the other subgroups. Fortunately, descriptive statistics that are sample moments can generally be obtained, so long as the analyst has access to all sample moments up to the order of the moment of interest.

---

[*] E-mail address: ben.oneill@anu.edu.au; ben.oneill@hotmail.com.
[**] Research School of Population Health, Australian National University, Canberra ACT 0200.

In this paper, we will examine some useful decompositions of sum-of-power statistics relating to corresponding decompositions of central moment statistics, to obtain descriptive sample moments of a pooled sample or a missing subgroup from the corresponding moments of the other groups. Our decompositions are implemented in the `sample.decomp` function in the `utilities` package in R (O'Neill 2021). We give an applied example showing how to use this function to obtain decomposition results for pooled samples or missing subgroups.

**1. Sum-of-powers decompositions up to fourth order**

Our analysis will allow decomposition to occur for any number of subgroups, but the simplest way to present this is to first consider the case of two subgroups that aggregate to give a single pooled sample. We will begin by considering two sample subgroups $\mathbf{x}_1 = (x_{1,1}, \ldots, x_{1,n_1})$ and $\mathbf{x}_2 = (x_{2,1}, \ldots, x_{2,n_2})$ and we denote their pooled sample by $\mathbf{x}_p = (x_1, \ldots, x_{n_1+n_2})$ with sample size decomposition $n_1 + n_2 = n$.[1] The sample means, sums-of-squares, sums-of-cubes and sums-of-quads are given using the following notation:

$$\bar{x}_1 = \frac{1}{n_1} \sum_{i=1}^{n_1} x_{1,i} \quad SS_1 = \sum_{i=1}^{n_1} (x_{1,i} - \bar{x}_1)^2 \quad SC_1 = \sum_{i=1}^{n_1} (x_{1,i} - \bar{x}_1)^3 \quad SQ_1 = \sum_{i=1}^{n_1} (x_{1,i} - \bar{x}_1)^4,$$

$$\bar{x}_2 = \frac{1}{n_2} \sum_{i=1}^{n_2} x_{2,i} \quad SS_2 = \sum_{i=1}^{n_2} (x_{2,i} - \bar{x}_2)^2 \quad SC_2 = \sum_{i=1}^{n_2} (x_{2,i} - \bar{x}_2)^3 \quad SQ_2 = \sum_{i=1}^{n_2} (x_{2,i} - \bar{x}_2)^4,$$

$$\bar{x}_p = \frac{1}{n_p} \sum_{i=1}^{n_p} x_{p,i} \quad SS_p = \sum_{i=1}^{n_p} (x_{p,i} - \bar{x}_p)^2 \quad SC_p = \sum_{i=1}^{n_p} (x_{p,i} - \bar{x}_p)^3 \quad SQ_p = \sum_{i=1}^{n_p} (x_{p,i} - \bar{x}_p)^4.$$

Given known sample sizes, the standard sample moments of interest for a dataset are functions of the quantities above. The sample variance is a scaled version of the sum-of-squares using Bessel's correction, and the sample skewness and sample kurtosis are then obtained from the sums-of-cubes and sums-of-quads by scaling with the sample variance (for some variations of the sample skewness and kurtosis, see Joanes and Gill 1998).

As a general rule, given knowledge of the sample sizes, we can decompose the pooled sums-of-powers into parts composed of sums-of-powers of the subgroups of equal and lower order.

---

[1] A simple correspondence between the elements of the pooled sample and the elements of the subgroups can be developed, but we will not need to be specific about that here.

Our first theorem shows decompositions for the sums-of-squares, sums-of-cubes, and sums-of-quads, allowing decomposition of quantities for a pooled sample or a missing subgroup.

**THEOREM 1 (Sample mean decompositions):** The sample means can be decomposed as:

$$\bar{x}_1 = \frac{n_\mathrm{p}\bar{x}_\mathrm{p} - n_2\bar{x}_2}{n_\mathrm{p} - n_2},$$

$$\bar{x}_2 = \frac{n_\mathrm{p}\bar{x}_\mathrm{p} - n_1\bar{x}_1}{n_\mathrm{p} - n_1},$$

$$\bar{x}_\mathrm{p} = \frac{n_1\bar{x}_1 + n_2\bar{x}_2}{n_1 + n_2}.$$

**THEOREM 2 (Sum-of-powers decompositions):** The sums-of-squares, sums-of-cubes and sums-of-quads can be decomposed as:

$$\mathrm{SS}_1 = \mathrm{SS}_\mathrm{p} - \mathrm{SS}_2 - \frac{n_2 n_\mathrm{p}}{n_\mathrm{p} - n_2} \cdot (\bar{x}_2 - \bar{x}_\mathrm{p})^2,$$

$$\mathrm{SS}_2 = \mathrm{SS}_\mathrm{p} - \mathrm{SS}_1 - \frac{n_1 n_\mathrm{p}}{n_\mathrm{p} - n_1} \cdot (\bar{x}_1 - \bar{x}_\mathrm{p})^2,$$

$$\mathrm{SS}_\mathrm{p} = \mathrm{SS}_1 + \mathrm{SS}_2 + \frac{n_1 n_2}{n_1 + n_2} \cdot (\bar{x}_1 - \bar{x}_2)^2,$$

$$\mathrm{SC}_1 = \mathrm{SC}_\mathrm{p} - \mathrm{SC}_2 - 3 \cdot \frac{n_\mathrm{p} \mathrm{SS}_2 - n_2 \mathrm{SS}_\mathrm{p}}{n_\mathrm{p} - n_2} \cdot (\bar{x}_2 - \bar{x}_\mathrm{p}) - \frac{n_\mathrm{p} n_2^2 + n_2 n_\mathrm{p}^2}{(n_\mathrm{p} - n_2)^2} \cdot (\bar{x}_2 - \bar{x}_\mathrm{p})^3,$$

$$\mathrm{SC}_2 = \mathrm{SC}_\mathrm{p} - \mathrm{SC}_1 - 3 \cdot \frac{n_\mathrm{p} \mathrm{SS}_1 - n_1 \mathrm{SS}_\mathrm{p}}{n_\mathrm{p} - n_1} \cdot (\bar{x}_1 - \bar{x}_\mathrm{p}) - \frac{n_\mathrm{p} n_1^2 + n_1 n_\mathrm{p}^2}{(n_\mathrm{p} - n_1)^2} \cdot (\bar{x}_1 - \bar{x}_\mathrm{p})^3,$$

$$\mathrm{SC}_\mathrm{p} = \mathrm{SC}_1 + \mathrm{SC}_2 + 3 \cdot \frac{n_2 \mathrm{SS}_1 - n_1 \mathrm{SS}_2}{n_1 + n_2} \cdot (\bar{x}_1 - \bar{x}_2) + \frac{n_1 n_2^3 - n_2 n_1^3}{(n_1 + n_2)^3} \cdot (\bar{x}_1 - \bar{x}_2)^3,$$

$$\mathrm{SQ}_1 = \mathrm{SQ}_\mathrm{p} - \mathrm{SQ}_2 - 4 \cdot \frac{n_\mathrm{p} \mathrm{SC}_2 - n_2 \mathrm{SC}_\mathrm{p}}{n_\mathrm{p} - n_2} \cdot (\bar{x}_2 - \bar{x}_\mathrm{p})$$
$$- 6 \cdot \frac{n_\mathrm{p}^2 \mathrm{SS}_2 - n_2^2 \mathrm{SS}_\mathrm{p}}{(n_\mathrm{p} - n_2)^2} \cdot (\bar{x}_2 - \bar{x}_\mathrm{p})^2 - \frac{n_2 n_\mathrm{p}^3 + n_2^2 n_\mathrm{p}^2 + n_2^3 n_\mathrm{p}}{(n_\mathrm{p} - n_2)^3} \cdot (\bar{x}_2 - \bar{x}_\mathrm{p})^4,$$

$$\mathrm{SQ}_2 = \mathrm{SQ}_\mathrm{p} - \mathrm{SQ}_1 - 4 \cdot \frac{n_\mathrm{p} \mathrm{SC}_1 - n_1 \mathrm{SC}_\mathrm{p}}{n_\mathrm{p} - n_1} \cdot (\bar{x}_1 - \bar{x}_\mathrm{p})$$
$$- 6 \cdot \frac{n_\mathrm{p}^2 \mathrm{SS}_1 - n_1^2 \mathrm{SS}_\mathrm{p}}{(n_\mathrm{p} - n_1)^2} \cdot (\bar{x}_1 - \bar{x}_\mathrm{p})^2 - \frac{n_1 n_\mathrm{p}^3 + n_1^2 n_\mathrm{p}^2 + n_1^3 n_\mathrm{p}}{(n_\mathrm{p} - n_1)^3} \cdot (\bar{x}_1 - \bar{x}_\mathrm{p})^4,$$

$$\mathrm{SQ}_\mathrm{p} = \mathrm{SQ}_1 + \mathrm{SQ}_2 + 4 \cdot \frac{n_2 \mathrm{SC}_1 - n_1 \mathrm{SC}_2}{n_1 + n_2} \cdot (\bar{x}_1 - \bar{x}_2)$$
$$+ 6 \cdot \frac{n_2^2 \mathrm{SS}_1 + n_1^2 \mathrm{SS}_2}{(n_1 + n_2)^2} \cdot (\bar{x}_1 - \bar{x}_2)^2 + \frac{n_1 n_2^4 + n_2 n_1^4}{(n_1 + n_2)^4} \cdot (\bar{x}_1 - \bar{x}_2)^4.$$

Decompositions for the sums-of-powers lead to corresponding decompositions for the sample central moments. The corresponding formulae are simple to derive but cumbersome to state, so we omit them in this paper. (Explicit decompositions for the pooled sample mean and sample variance are shown in O'Neill 2014, p. 283, Result 1.) The decompositions are also useful because they give rise to iterative formulae for updating the sums-of-powers one data point at a time. To do this, we now suppose that we have an initial dataset $\mathbf{x}_n = (x_1, \ldots, x_n)$ consisting of $n$ data points and we observe a new value $x_{n+1} = x$ (which we will treat as our second subgroup). In this context, we now denote all sample moments for the initial sample using a subscript $n$ to denote the number of data points in that initial sample. We then have the following iterative formulae.

**THEOREM 3 (Iterative updating of sums-of-powers):** The sample mean and sums-of-powers statistics with $n + 1$ data points can be written in terms of the sample mean and sums-of-powers statistics with $n$ data points and the new value $x$ as follows:[2]

$$\bar{x}_{n+1} = \frac{n\bar{x}_n + x}{n+1},$$

$$\mathrm{SS}_{n+1} = \mathrm{SS}_n + \frac{n}{n+1} \cdot (\bar{x}_n - x)^2,$$

$$\mathrm{SC}_{n+1} = \mathrm{SC}_n + \frac{3 \cdot \mathrm{SS}_n}{n+1} \cdot (\bar{x}_n - x) - \frac{n(n-1)}{(n+1)^2} \cdot (\bar{x}_n - x)^3,$$

$$\mathrm{SQ}_{n+1} = \mathrm{SQ}_n + \frac{4 \cdot \mathrm{SC}_n}{n+1} \cdot (\bar{x}_n - x) + \frac{6 \cdot \mathrm{SS}_n}{(n+1)^2} \cdot (\bar{x}_n - x)^2 + \frac{n(1+n^3)}{(n+1)^4} \cdot (\bar{x}_n - x)^4.$$

Iterative formulae for the sums-of-powers are useful for computational purposes, since they avoid arithmetic underflow problems that can arise when computing using the base formula (see e.g., Ling 1974; Chan, Golub and LeVeque 1983; Phien 1988). These formulae are useful when programming functions to compute the sample central moments, particularly for high-order moments. For example, the iterative formulae in Theorem 3 are used in the **utilities** package in **R** (O'Neill 2021) for the sample moment functions **skewness**, **kurtosis** and **moments**. Iterative computation using these formulae is generally slower but more stable than using direct "one-step" formulae for the sample central moments.

---

[2] Note the notational change in this theorem. In this context the subscripts on the moments now refers to the number of data points in the sample under consideration.

## 2. Extensions to allow more subgroups and higher powers

The above decomposition formulae are for pooled samples composed of two subgroups. For pooled samples composed of more than two subgroups we can apply the formulae iteratively to obtain a desired sum-of-powers quantity. This iterative method has the same benefit it has when iteratively computing moments from a single sample — i.e., iterative computation reduces the risk of underflow problems in the computation. Nevertheless, if one is confident that underflow problems are not at issue, it is useful to examine the extended case explicitly, looking at formulae that allow direct pooling for an arbitrary number of groups.

We now consider a problem with $k$ subgroups with size decomposition $n_1 + \cdots + n_k = n$. In order to extend our analysis to this larger class of subgroups, we now replace the subscript notation for our pooled sample with explicit notation $1{:}k$ for the pooling of subgroups over all subgroups. We begin by looking at a formula for the pooled sums-of-powers for an arbitrary number of subgroups, written in terms of the subgroup sums-of-powers. To do this, we define the general sum-of-powers statistics:

$$\text{SP}_{1:k}^p = \sum_{i=1}^{n}(x_i - \bar{x}_{1:k})^p \qquad \bar{x}_{1:k} = \frac{1}{n}\sum_{i=1}^{n}x_i,$$

and we define the corresponding subgroup quantities as:

$$\text{SP}_{\ell}^p = \sum_{i=1}^{n_\ell}(x_{\ell,i} - \bar{x}_\ell)^p \qquad \bar{x}_\ell = \frac{1}{n_\ell}\sum_{i=1}^{n_\ell}x_{\ell,i}.$$

(Note here that we have used a dual notation for the $n$ data points; when considered as part of the pooled sample we denote them as $x_1, \ldots, x_n$, but when considered as part of the subgroups we denote them as $x_{1,1}, \ldots, x_{1,n_1}, \ldots, x_{k,1}, \ldots, x_{1,n_k}$. It is not necessary to specify the specific correspondence here, since we are only ever summing over subgroups or the pooled sample.) The values $p = 2,3,4$ give sums-of-squares, sums-of-cubes and sums-of-quads respectively.

**THEOREM 4 (Generalised sum-of-powers decomposition):** The pooled sum-of-powers for an arbitrary number of subgroups can be decomposed as:

$$\text{SP}_{1:k}^p = \sum_{s=0}^{p}\binom{p}{s}\sum_{\ell=1}^{k}\text{SP}_\ell^{p-s}\left(\bar{x}_\ell - \frac{1}{n}\sum_{i=1}^{k}n_i\bar{x}_i\right)^s.$$

Theorem 4 gives us a generalised decomposition formula for the pooled sum-of-powers for any number of subgroups. This formula can be used as a basis to derive specific decompositions for particular powers or particular numbers of subgroups. It is useful to give specific results for the pooled sum-of-squares $\text{SS}_{1:k}$, the pooled sum-of-cubes $\text{SC}_{1:k}$ and the pooled sum-of-quads $\text{SQ}_{1:k}$ using all $k$ subgroups. In Theorem 5 we give forms for the pooled sum-of-squares, sum-of-cubes and sum-of-quads using relevant subgroup quantities.

**THEOREM 5:** The pooled sum-of-squares, sum-of-cubes and sum-of-quads can be decomposed in terms of the subgroup statistics as:

$$\text{SS}_{1:k} = \text{SS}_1 + \text{SS}_2 + \cdots + \text{SS}_k + \sum_{i=1}^{k} n_i \bar{x}_i^2 - \frac{1}{n_1 + \cdots + n_k} \cdot \left( \sum_{i=1}^{k} n_i \bar{x}_i \right)^2,$$

$$\text{SC}_{1:k} = \text{SC}_1 + \text{SC}_2 + \cdots + \text{SC}_k + 3 \sum_{\ell=1}^{k} \text{SS}_\ell \left( \bar{x}_\ell - \frac{1}{n_1 + \cdots + n_k} \sum_{i=1}^{k} n_i \bar{x}_i \right)$$

$$+ \sum_{\ell=1}^{k} n_\ell \left( \bar{x}_\ell - \frac{1}{n_1 + \cdots + n_k} \sum_{i=1}^{k} n_i \bar{x}_i \right)^3,$$

$$\text{SQ}_{1:k} = \text{SQ}_1 + \text{SQ}_2 + \cdots + \text{SQ}_k + 4 \sum_{\ell=1}^{k} \text{SC}_\ell \left( \bar{x}_\ell - \frac{1}{n_1 + \cdots + n_k} \sum_{i=1}^{k} n_i \bar{x}_i \right)$$

$$+ 6 \sum_{\ell=1}^{k} \text{SS}_\ell \left( \bar{x}_\ell - \frac{1}{n_1 + \cdots + n_k} \sum_{i=1}^{k} n_i \bar{x}_i \right)^2$$

$$+ \sum_{\ell=1}^{k} n_\ell \left( \bar{x}_\ell - \frac{1}{n_1 + \cdots + n_k} \sum_{i=1}^{k} n_i \bar{x}_i \right)^4.$$

The formulae in Theorem 5 can be used to obtained pooled sample central moments from the sample central moments of a set of underlying subgroups. This is a "one-step" process that is an alternative to iterative updating using the formulae in Theorem 2. Corresponding "one-step" formulae for the missing subgroup sum-of-squares, sum-of-cubes and sum-of-quads can be obtained by combining formulae in Theorems 2 and 5. However, in practice, it is simplest to obtain sum-of-powers quantities for a missing subgroup by first using Theorem 5 to compute the pooled sums-of-powers for all the non-missing subgroups (as intermediate quantities) and then using the formulae in Theorem 2 to obtain the sums-of-powers for the missing subgroup.

# 3. Implementation in the `utilities` package in `R`

The decomposition results in this paper are implemented in `R` in the `utilities` package (O'Neill 2021) to obtain decompositions of the sample central moments up to fourth order. Users can use the `sample.decomp` function to compute the sample moment statistics for a pooled sample or a missing group given the corresponding sample moments from the remaining groups. The arguments for this function are shown in Table 1 below. The user can either input vectors containing the sample sizes and sample moments for each group (shown in yellow) or they can input a full moments object produced by the `moments` function (shown in orange). Variations of the sample skewness and sample kurtosis are specified with appropriate character inputs (shown in blue) and the names of the groups can also be specified (shown in green). By default, the function treats the input groups as subgroups and computes the statistics for the pooled sample from these subgroups. However, the user can specify that one of the groups is the pooled group (shown in purple), in which case the function treats this group as the pooled sample and computes the statistics for the "other" subgroup that makes up the pooled sample.

| Argument | Type | Description |
|---|---|---|
| `moments` | Moments | A moments object (produced by the `moments` function) containing the sample sizes and sample moments of all groups (mean, variance, skewness, kurtosis) |
| `n` | Integer | A vector containing the sample sizes of the groups |
| `sample.mean`† | Numeric | A vector containing the sample means of the groups |
| `sample.sd`† | Numeric | A vector containing the sample standard deviation of the groups |
| `sample.var`† | Numeric | A vector containing the sample variance of the groups |
| `sample.skew`† | Numeric | A vector containing the sample skewness of the groups |
| `sample.kurt`† | Numeric | A vector containing the sample kurtosis of the groups |
| `names`* | Character | A vector of names for the groups |
| `skew.type`* | Character | A character specifying the type of sample skewness statistic being used (including `'Moment'`, `'Fisher Pearson'`, `'Adjusted Fisher Pearson'` and various other options specifying statistical software) |
| `kurt.type`* | Character | A character specifying the type of sample kurtosis statistic being used (including `'Moment'`, `'Fisher Pearson'`, `'Adjusted Fisher Pearson'` and various other options specifying statistical software) |
| `kurt.excess`* | Logical | If `TRUE` the sample kurtosis is excess kurtosis rather than raw kurtosis |
| `pooled` | Integer | An integer specifying the group that is the pooled sample (optional) If no input is given then none of the inputs are treated as the pooled sample |
| `include.sd`* | Logical | If `TRUE` the output includes a column of values for the standard deviation |
| † All sample moment statistics are optional — the function will give output up to the highest sample moment for which all sample moments up to that order are supplied for the other groups. <br> * Default values are given in the function. | | |

**Table 1:** Arguments for the `sample.decomp` function

In the code below we show the sample decomposition for a simple case with three subgroups, where we wish to compute the sample moment statistics for the pooled sample. We use the **sample.decomp** function to compute the sample moment statistics for the pooled sample from the sample moment statistics of the three subgroups, without use of the datasets. (The commands are shown in black and output from **R** is shown in blue.) It is simple to confirm that the statistics computed for the pooled sample match the values computed from the underlying pooled dataset.[3]

```
#Create some subgroups of mock data and a pooled dataset
set.seed(1)
N    <- c(28, 44, 51)
SUB1 <- rnorm(N[1])
SUB2 <- rnorm(N[2])
SUB3 <- rnorm(N[3])
POOL <- c(SUB1, SUB2, SUB3)

#Compute sample statistics for subgroups
library(utilities)
MEAN <- c(     mean(SUB1),     mean(SUB2),     mean(SUB3))
VAR  <- c(      var(SUB1),      var(SUB2),      var(SUB3))
SKEW <- c(skewness(SUB1), skewness(SUB2), skewness(SUB3))
KURT <- c(kurtosis(SUB1), kurtosis(SUB2), kurtosis(SUB3))

#Compute sample decomposition
sample.decomp(n = N,
              sample.mean = MEAN,
              sample.var  = VAR,
              sample.skew = SKEW,
              sample.kurt = KURT,
              skew.type   = 'Fisher Pearson',
              kurt.type   = 'Fisher Pearson',
              kurt.excess = FALSE,
              include.sd  = FALSE)

            n sample.mean sample.var sample.skew sample.kurt
1          28  0.09049834  0.9013829 -0.76480085    3.174128
2          44  0.18637936  0.8246700  0.36539179    3.112901
3          51  0.05986594  0.6856030  0.30762810    2.306243
--pooled-- 123 0.11209600  0.7743711  0.04697463    2.951960
```

The function includes an option **pooled** to specify that one of the groups is the pooled group. In this case the function computes the sample moment statistics for the remaining subgroup that is required to aggregate to give the specified pooled sample. In the code below we show the sample decomposition obtained by specifying the moments for the first two subgroups and

---

[3] Note that the **utilities** package also contains the function **moments**, which allows the user to create a table of moments from a list of sample groups. The output table from this function can be used directly as an input into the **sample.decomp** function. We have not used this method here, since we wish to show clearly that the latter function computes the pooled moments without access to the original data.

the pooled sample. In this case the function outputs the moments of the "other" subgroup. As can be seen from the output, this replicates the same sample moment statistics as the previous application of the function. (The output is slightly different because we now label the missing subgroup as the "other" group. Aside from this cosmetic difference, the output is the same to within tiny margins of arithmetic error.[4])

```
#Compute sample decomposition
N.P    <- length(POOL)
MEAN.P <- mean(POOL)
VAR.P  <- var(POOL)
SKEW.P <- skewness(POOL)
KURT.P <- kurtosis(POOL)

sample.decomp(n = c(N[1:2], N.P),
              sample.mean = c(MEAN[1:2], MEAN.P),
              sample.var  = c( VAR[1:2],  VAR.P),
              sample.skew = c(SKEW[1:2], SKEW.P),
              sample.kurt = c(KURT[1:2], KURT.P),
              skew.type   = 'Fisher Pearson',
              kurt.type   = 'Fisher Pearson',
              kurt.excess = FALSE,
              pooled      = 3,
              include.sd  = FALSE)

             n sample.mean sample.var sample.skew sample.kurt
1           28  0.09049834  0.9013829 -0.76480085    3.174128
2           44  0.18637936  0.8246700  0.36539179    3.112901
--other--   51  0.05986594  0.6856030  0.30762810    2.306243
--pooled-- 123  0.11209600  0.7743711  0.04697463    2.951960
```

It is again important to note that it is only possible to compute the decompositions for sample moment statistics if the user has knowledge of all sample moments from the other groups up to the desired order. In the event that any of the sample moments for the other groups are missing it becomes impossible not only to compute decompositions for that sample moment, but also for all sample moments of higher order. In the **sample.decomp** function the output only goes up to the highest order for which all moments are provided for the input groups.

### 4. Concluding remarks

Sample moment decompositions are useful in allowing analysts to deal with cases where they wish to aggregate sample subgroups with information on their sample moments but no access to the underlying data. The decompositions are also useful because they give iterative formulae for updating the sample moments one data point at a time using previously computed moments (i.e., without recomputing the moments from all data points). Formulae exist to compute the sample moments of a pooled sample from the sample moments of its subgroups, or to compute the sample moments of a single missing subgroup from the sample moments of the pooled sample and the other subgroups. These are useful in cases where an analyst wishes to aggregate multiple subgroups or find a missing subgroup from an aggregation, in cases where there is no access to the underlying data in the subgroups or pooled sample.

---

[4] There is a tiny amount of arithmetic error in the computations due to rounding. Comparing the two tables we get a maximum disparity of $1.332268 \times 10^{-15}$ in the computations.

This paper has provided decomposition formulae for the underlying sum-of-powers quantities used in these problems. These decompositions are implemented in the `sample.decomp` function in the `utilities` package in `R`. This function is simple and user-friendly; it allows the user to input vectors of sample moments for the groups for which such information is available, and it gives a simple output that shows the sample moment statistics for all groups. This adds to the arsenal of tools for practicing statistical analysts, by allowing easy recovery of moments for a pooled sample or a missing subgroup when underlying data is not available.

# Appendix: Proof of Theorems

**PROOF OF THEOREM 1:** These results all follow easily from the fact that $n_p \bar{x}_p = n_1 \bar{x}_1 + n_2 \bar{x}_2$ and $n_p = n_1 + n_2$. ∎

For convenience, we will split the proof of Theorem 2 into parts (a), (b) and (c) corresponding to the decompositions for the sums-of-squares, sums-of-cubes, and sums-of-quads. We will prove each of these in order, appealing to previously proved results as we go.

**PROOF OF THEOREM 2(A):** With a little algebra we can obtain the preliminary result:

$$(\bar{x}_1 - \bar{x}_p)^k = \left(\frac{n_2}{n_1 + n_2}\right)^k (\bar{x}_1 - \bar{x}_2)^k,$$

$$(\bar{x}_2 - \bar{x}_p)^k = \left(\frac{n_1}{n_1 + n_2}\right)^k (\bar{x}_2 - \bar{x}_1)^k,$$

$$(\bar{x}_1 - \bar{x}_p)^k = (-1)^k \left(\frac{n_2}{n_1}\right)^k (\bar{x}_2 - \bar{x}_p)^k.$$

Using these results, the pooled sum-of-squares can be decomposed as:

$$\begin{aligned}
SS_p &\equiv \sum_{i=1}^{n_p} (x_i - \bar{x}_p)^2 \\
&= \sum_{i=1}^{n_1} (x_i - \bar{x}_p)^2 + \sum_{i=1}^{n_2} (x_{n_1+i} - \bar{x}_p)^2 \\
&= \sum_{i=1}^{n_1} (x_i - \bar{x}_1 + \bar{x}_1 - \bar{x}_p)^2 + \sum_{i=1}^{n_2} (x_{n_1+i} - \bar{x}_2 + \bar{x}_2 - \bar{x}_p)^2 \\
&= \sum_{i=1}^{n_1} (x_i - \bar{x}_1)^2 + \sum_{i=1}^{n_2} (x_{n_1+i} - \bar{x}_1)^2 + n_1(\bar{x}_1 - \bar{x}_p)^2 + n_2(\bar{x}_2 - \bar{x}_p)^2 \\
&= SS_1 + SS_2 + n_1(\bar{x}_1 - \bar{x}_p)^2 + n_2(\bar{x}_2 - \bar{x}_p)^2 \\
&= SS_1 + SS_2 + n_1 \left(\frac{n_2}{n_1 + n_2}\right)^2 (\bar{x}_1 - \bar{x}_2)^2 + n_2 \left(\frac{n_1}{n_1 + n_2}\right)^2 (\bar{x}_1 - \bar{x}_2)^2 \\
&= SS_1 + SS_2 + \frac{n_1 n_2}{n_1 + n_2} \cdot (\bar{x}_1 - \bar{x}_2)^2.
\end{aligned}$$

In order to get the decompositions for the individual samples we first establish that:

$$\mathrm{SS_p} = \mathrm{SS}_1 + \mathrm{SS}_2 + n_1(\bar{x}_1 - \bar{x}_\mathrm{p})^2 + n_2(\bar{x}_2 - \bar{x}_\mathrm{p})^2$$

$$= \mathrm{SS}_1 + \mathrm{SS}_2 + n_1 \left(\frac{n_2}{n_1}\right)^2 (\bar{x}_2 - \bar{x}_\mathrm{p})^2 + n_2(\bar{x}_2 - \bar{x}_\mathrm{p})^2$$

$$= \mathrm{SS}_1 + \mathrm{SS}_2 + \frac{n_2(n_1 + n_2)}{n_1} \cdot (\bar{x}_2 - \bar{x}_\mathrm{p})^2$$

$$= \mathrm{SS}_1 + \mathrm{SS}_2 + \frac{n_2 n_\mathrm{p}}{n_\mathrm{p} - n_2} \cdot (\bar{x}_2 - \bar{x}_\mathrm{p})^2,$$

A simple rearrangement gives the decomposition for the sum-of-squares $\mathrm{SS}_1$ and an analogous argument gives the corresponding decomposition for the sum-of-squares $\mathrm{SS}_2$. The establishes the decompositions for the sums-of-squares. ∎

**PROOF OF THEOREM 2(B):** The pooled sum-of-cubes can be decomposed as:

$$\mathrm{SC_p} \equiv \sum_{i=1}^{n_\mathrm{p}} (x_i - \bar{x}_\mathrm{p})^3$$

$$= \sum_{i=1}^{n_1} (x_i - \bar{x}_\mathrm{p})^3 + \sum_{i=1}^{n_2} (x_{n_1+i} - \bar{x}_\mathrm{p})^3$$

$$= \sum_{i=1}^{n_1} (x_i - \bar{x}_1 + \bar{x}_1 - \bar{x}_\mathrm{p})^3 + \sum_{i=1}^{n_2} (x_{n_1+i} - \bar{x}_2 + \bar{x}_2 - \bar{x}_\mathrm{p})^3$$

$$= \sum_{i=1}^{n_1} (x_i - \bar{x}_1)^3 + 3(\bar{x}_1 - \bar{x}_\mathrm{p}) \sum_{i=1}^{n_1} (x_i - \bar{x}_1)^2 + n_1(\bar{x}_1 - \bar{x}_\mathrm{p})^3$$

$$+ \sum_{i=1}^{n_2} (x_{n_1+i} - \bar{x}_2)^3 + 3(\bar{x}_2 - \bar{x}_\mathrm{p}) \sum_{i=1}^{n_2} (x_{n_1+i} - \bar{x}_2)^2 + n_2(\bar{x}_2 - \bar{x}_\mathrm{p})^3$$

$$= \mathrm{SC}_1 + \mathrm{SC}_2 + 3(\bar{x}_1 - \bar{x}_\mathrm{p})\mathrm{SS}_1 + 3(\bar{x}_2 - \bar{x}_\mathrm{p})\mathrm{SS}_2 + n_1(\bar{x}_1 - \bar{x}_\mathrm{p})^3 + n_2(\bar{x}_2 - \bar{x}_\mathrm{p})^3$$

$$= \mathrm{SC}_1 + \mathrm{SC}_2 + 3\left(\frac{n_2}{n_1 + n_2}\right)(\bar{x}_1 - \bar{x}_2)\mathrm{SS}_1 + 3\left(\frac{n_1}{n_1 + n_2}\right)(\bar{x}_2 - \bar{x}_1)\mathrm{SS}_2$$

$$+ n_1 \left(\frac{n_2}{n_1 + n_2}\right)^3 (\bar{x}_1 - \bar{x}_2)^3 + n_2 \left(\frac{n_1}{n_1 + n_2}\right)^3 (\bar{x}_2 - \bar{x}_1)^3$$

$$= \mathrm{SC}_1 + \mathrm{SC}_2 + \frac{3 n_2 \mathrm{SS}_1}{n_1 + n_2} (\bar{x}_1 - \bar{x}_2) + \frac{3 n_1 \mathrm{SS}_2}{n_1 + n_2} (\bar{x}_2 - \bar{x}_1)$$

$$+ \frac{n_1 n_2^3}{(n_1 + n_2)^3} (\bar{x}_1 - \bar{x}_2)^3 + \frac{n_1^3 n_2}{(n_1 + n_2)^3} (\bar{x}_2 - \bar{x}_1)^3$$

$$= \mathrm{SC}_1 + \mathrm{SC}_2 + 3 \cdot \frac{n_2 \mathrm{SS}_1 - n_1 \mathrm{SS}_2}{n_1 + n_2} \cdot (\bar{x}_1 - \bar{x}_2) + \frac{n_1 n_2}{n_1 + n_2} \cdot \frac{n_2^2 - n_1^2}{(n_1 + n_2)^2} \cdot (\bar{x}_1 - \bar{x}_2)^3.$$

In order to get the decompositions for the individual samples we first establish that:

$$SS_2 - \frac{n_2}{n_1}SS_1 = SS_2 - \frac{n_2}{n_1}\left[SS_p - SS_2 - \frac{n_2 n_p}{n_p - n_2} \cdot (\bar{x}_2 - \bar{x}_p)^2\right]$$

$$= \frac{n_1 + n_2}{n_1} \cdot SS_2 - \frac{n_2}{n_1} \cdot SS_p + \frac{n_2}{n_1} \cdot \frac{n_2 n_p}{n_p - n_2} \cdot (\bar{x}_2 - \bar{x}_p)^2$$

$$= \frac{n_p}{n_p - n_2} \cdot SS_2 - \frac{n_2}{n_p - n_2} \cdot SS_p + \frac{n_2^2 n_p}{(n_p - n_2)^2} \cdot (\bar{x}_2 - \bar{x}_p)^2$$

$$= \frac{1}{n_p - n_2}\left[n_p SS_2 - n_2 SS_p + \frac{n_2^2 n_p}{n_p - n_2} \cdot (\bar{x}_2 - \bar{x}_p)^2\right].$$

We then have:

$$SC_p = SC_1 + SC_2 + 3(\bar{x}_1 - \bar{x}_p)SS_1 + 3(\bar{x}_2 - \bar{x}_p)SS_2 + n_1(\bar{x}_1 - \bar{x}_p)^3 + n_2(\bar{x}_2 - \bar{x}_p)^3$$

$$= SC_1 + SC_2 + 3\left(SS_2 - \frac{n_2}{n_1}SS_1\right)(\bar{x}_2 - \bar{x}_p) + \left(n_2 - n_1\left(\frac{n_2}{n_1}\right)^3\right)(\bar{x}_2 - \bar{x}_p)^3$$

$$= SC_1 + SC_2 + 3\left(SS_2 - \frac{n_2}{n_1}SS_1\right)(\bar{x}_2 - \bar{x}_p) + n_2 \cdot \frac{n_1^2 - n_2^2}{n_1^2} \cdot (\bar{x}_2 - \bar{x}_p)^3$$

$$= SC_1 + SC_2 + \frac{3}{n_p - n_2}\left[n_p SS_2 - n_2 SS_p + \frac{n_2^2 n_p}{n_p - n_2} \cdot (\bar{x}_2 - \bar{x}_p)^2\right](\bar{x}_2 - \bar{x}_p)$$

$$+ n_2 \cdot \frac{(n_p - n_2)^2 - n_2^2}{(n_p - n_2)^2} \cdot (\bar{x}_2 - \bar{x}_p)^3.$$

$$= SC_1 + SC_2 + 3 \cdot \frac{n_p SS_2 - n_2 SS_p}{n_p - n_2} \cdot (\bar{x}_2 - \bar{x}_p)$$

$$+ 3 \cdot \frac{n_2^2 n_p}{(n_p - n_2)^2} \cdot (\bar{x}_2 - \bar{x}_p)^3 + \frac{n_2 n_p^2 - 2 n_p n_2^2}{(n_p - n_2)^2} \cdot (\bar{x}_2 - \bar{x}_p)^3$$

$$= SC_1 + SC_2 + 3 \cdot \frac{n_p SS_2 - n_2 SS_p}{n_p - n_2} \cdot (\bar{x}_2 - \bar{x}_p) + \frac{n_2 n_p}{n_p - n_2} \cdot \frac{n_2 + n_p}{n_p - n_2} \cdot (\bar{x}_2 - \bar{x}_p)^3.$$

A simple rearrangement gives the decomposition for the sum-of-cubes $SC_1$ and an analogous argument gives the corresponding decomposition for the sum-of-cubes $SC_2$. The establishes the decompositions for the sums-of-cubes. ∎

**PROOF OF THEOREM 2©:** The pooled sum-of-quads can be decomposed as:

$$SC_p \equiv \sum_{i=1}^{n_p}(x_i - \bar{x}_p)^4$$

$$= \sum_{i=1}^{n_1}(x_i - \bar{x}_p)^4 + \sum_{i=1}^{n_2}(x_{n_1+i} - \bar{x}_p)^4$$

$$= \sum_{i=1}^{n_1}(x_i - \bar{x}_1 + \bar{x}_1 - \bar{x}_p)^4 + \sum_{i=1}^{n_2}(x_{n_1+i} - \bar{x}_2 + \bar{x}_2 - \bar{x}_p)^4$$

$$= \sum_{i=1}^{n_1}(x_i - \bar{x}_1)^4 + 4(\bar{x}_1 - \bar{x}_p)\sum_{i=1}^{n_1}(x_i - \bar{x}_1)^3 + 6(\bar{x}_1 - \bar{x}_p)^2\sum_{i=1}^{n_1}(x_i - \bar{x}_1)^2$$

$$+ \sum_{i=1}^{n_2}(x_{n_1+i} - \bar{x}_2)^4 + 4(\bar{x}_2 - \bar{x}_p)\sum_{i=1}^{n_2}(x_{n_1+i} - \bar{x}_2)^3 + 6(\bar{x}_2 - \bar{x}_p)^2\sum_{i=1}^{n_2}(x_{n_1+i} - \bar{x}_2)^2$$

$$+ n_1(\bar{x}_1 - \bar{x}_p)^4 + n_2(\bar{x}_2 - \bar{x}_p)^3$$

$$= SQ_1 + SQ_2 + 4(\bar{x}_1 - \bar{x}_p)SC_1 + 4(\bar{x}_2 - \bar{x}_p)SC_2$$

$$+ 6(\bar{x}_1 - \bar{x}_p)^2 SS_1 + 6(\bar{x}_2 - \bar{x}_p)^2 SS_2 + n_1(\bar{x}_1 - \bar{x}_p)^4 + n_2(\bar{x}_2 - \bar{x}_p)^4$$

$$= SQ_1 + SQ_2 + 4\left(\frac{n_2}{n_1+n_2}\right)(\bar{x}_1 - \bar{x}_2)SC_1 - 4\left(\frac{n_1}{n_1+n_2}\right)(\bar{x}_1 - \bar{x}_2)SC_2$$

$$+ 6\left(\frac{n_2}{n_1+n_2}\right)^2(\bar{x}_1 - \bar{x}_2)^2 SS_1 + 6\left(\frac{n_1}{n_1+n_2}\right)^2(\bar{x}_1 - \bar{x}_2)^2 SS_2$$

$$+ n_1\left(\frac{n_2}{n_1+n_2}\right)^4(\bar{x}_1 - \bar{x}_2)^4 + n_2\left(\frac{n_1}{n_1+n_2}\right)^4(\bar{x}_1 - \bar{x}_2)^4$$

$$= SQ_1 + SQ_2 + 4 \cdot \frac{n_2 SC_1 - n_1 SC_2}{n_1 + n_2} \cdot (\bar{x}_1 - \bar{x}_2)$$

$$+ 6 \cdot \frac{n_2^2 SS_1 + n_1^2 SS_2}{(n_1+n_2)^2} \cdot (\bar{x}_1 - \bar{x}_2)^2 + \frac{n_1 n_2^4 + n_2 n_1^4}{(n_1+n_2)^4}(\bar{x}_1 - \bar{x}_2)^4.$$

In order to get the decompositions for the individual samples we first establish that:

$$SC_2 - \frac{n_2}{n_1}SC_1 = SC_2 - \frac{n_2}{n_1} \cdot SC_p + \frac{n_2}{n_1} \cdot SC_2$$

$$+ 3 \cdot \frac{n_2}{n_1} \cdot \frac{n_p SS_2 - n_2 SS_p}{n_p - n_2} \cdot (\bar{x}_2 - \bar{x}_p) + \frac{n_2}{n_1} \cdot \frac{n_p n_2^2 + n_2 n_p^2}{(n_p - n_2)^2} \cdot (\bar{x}_2 - \bar{x}_p)^3$$

$$= \frac{n_p SC_2 - n_2 SC_p}{n_p - n_2} + 3 \cdot \frac{n_2 n_p SS_2 - n_2^2 SS_p}{(n_p - n_2)^2} \cdot (\bar{x}_2 - \bar{x}_p) + \frac{n_p n_2^3 + n_2^2 n_p^2}{(n_p - n_2)^3} \cdot (\bar{x}_2 - \bar{x}_p)^3$$

and:

$$SS_2 + \left(\frac{n_2}{n_1}\right)^2 SS_1 = SS_2 + \left(\frac{n_2}{n_1}\right)^2 SS_p - \left(\frac{n_2}{n_1}\right)^2 SS_2 - \left(\frac{n_2}{n_1}\right)^2 \frac{n_2 n_p}{n_p - n_2} \cdot (\bar{x}_2 - \bar{x}_p)^2$$

$$= \frac{n_p(n_p - 2n_2)SS_2 + n_2^2 SS_p}{(n_p - n_2)^2} - \frac{n_2^3 n_p}{(n_p - n_2)^3} \cdot (\bar{x}_2 - \bar{x}_p)^2$$

We then have:

$$SQ_p = SQ_1 + SQ_2 + 4(\bar{x}_1 - \bar{x}_p)SC_1 + 4(\bar{x}_2 - \bar{x}_p)SC_2$$

$$+ 6(\bar{x}_1 - \bar{x}_p)^2 SS_1 + 6(\bar{x}_2 - \bar{x}_p)^2 SS_2 + n_1(\bar{x}_1 - \bar{x}_p)^4 + n_2(\bar{x}_2 - \bar{x}_p)^4$$

$$\begin{aligned}
&= SQ_1 + SQ_2 - 4\frac{n_2}{n_1}(\bar{x}_2 - \bar{x}_p)SC_1 + 4(\bar{x}_2 - \bar{x}_p)SC_2 \\
&\quad + 6\left(\frac{n_2}{n_1}\right)^2 (\bar{x}_2 - \bar{x}_p)^2 SS_1 + 6(\bar{x}_2 - \bar{x}_p)^2 SS_2 \\
&\quad + n_1\left(\frac{n_2}{n_1}\right)^4 (\bar{x}_2 - \bar{x}_p)^4 + n_2(\bar{x}_2 - \bar{x}_p)^4 \\
&= SQ_1 + SQ_2 + 4\left(SC_2 - \frac{n_2}{n_1}SC_1\right)(\bar{x}_2 - \bar{x}_p) \\
&\quad + 6\left(SS_2 + \left(\frac{n_2}{n_1}\right)^2 SS_1\right)(\bar{x}_2 - \bar{x}_p)^2 + \left(n_2 + \left(\frac{n_2}{n_1}\right)^4 n_1\right)(\bar{x}_2 - \bar{x}_p)^4 \\
&= SQ_1 + SQ_2 + 4 \cdot \frac{n_p SC_2 - n_2 SC_p}{n_p - n_2} \cdot (\bar{x}_2 - \bar{x}_p) \\
&\quad + 12 \cdot \frac{n_2 n_p SS_2 - n_2^2 SS_p}{(n_p - n_2)^2} \cdot (\bar{x}_2 - \bar{x}_p)^2 + 4 \cdot \frac{n_p n_2^3 + n_2^2 n_p^2}{(n_p - n_2)^3} \cdot (\bar{x}_2 - \bar{x}_p)^4 \\
&\quad + 6 \cdot \frac{n_p(n_p - 2n_2)SS_2 + n_2^2 SS_p}{(n_p - n_2)^2} \cdot (\bar{x}_2 - \bar{x}_p)^2 - 6 \cdot \frac{n_2^3 n_p}{(n_p - n_2)^3} \cdot (\bar{x}_2 - \bar{x}_p)^4 \\
&\quad + \frac{n_2 n_p^3 - 3n_2^2 n_p^2 + 3n_2^3 n_p}{(n_p - n_2)^3} \cdot (\bar{x}_2 - \bar{x}_p)^4 \\
&= SQ_1 + SQ_2 + 4 \cdot \frac{n_p SC_2 - n_2 SC_p}{n_p - n_2} \cdot (\bar{x}_2 - \bar{x}_p) \\
&\quad + 6 \cdot \frac{n_p^2 SS_2 - n_2^2 SS_p}{(n_p - n_2)^2} \cdot (\bar{x}_2 - \bar{x}_p)^2 + \frac{n_2 n_p^3 + n_2^2 n_p^2 + n_2^3 n_p}{(n_p - n_2)^3} \cdot (\bar{x}_2 - \bar{x}_p)^4.
\end{aligned}$$

A simple rearrangement gives the decomposition for the sum-of-quads $SQ_1$ and an analogous argument gives the corresponding decomposition for the sum-of-quads $SQ_2$. The establishes the decompositions for the sums-of-cubes. ∎

**PROOF OF THEOREM 3:** These results follow easily using substitution into the formulae for the pooled sample statistics in Theorems 1-2, where the first sample is $\mathbf{x} = (x_1, \ldots, x_n)$ and the second sample is the single data point $x$. For this substitution we have:

$$\begin{aligned}
n_1 &= n & n_2 &= 1, \\
\bar{x}_1 &= \bar{x}_n & \bar{x}_2 &= x, \\
SS_1 &= SS_n & SS_2 &= 0, \\
SC_1 &= SC_n & SC_2 &= 0, \\
SQ_1 &= SQ_n & SQ_2 &= 0.
\end{aligned}$$

Substitution of these values gives the desired results. ∎

**PROOF OF THEOREM 4:** We have:

$$\text{SP}_{1:k}^p = \sum_{i=1}^n (x_i - \bar{x}_{1:k})^p = \sum_{\ell=1}^k \sum_{i=1}^{n_\ell} (x_{\ell,i} - \bar{x}_{1:k})^p$$

$$= \sum_{\ell=1}^k \sum_{i=1}^{n_\ell} (x_{\ell,i} - \bar{x}_\ell + \bar{x}_\ell - \bar{x}_{1:k})^p$$

$$= \sum_{\ell=1}^k \left[ \sum_{s=0}^p \binom{p}{s} \left( \sum_{i=1}^{n_\ell} (x_{\ell,i} - \bar{x}_\ell)^{p-s} \right) (\bar{x}_\ell - \bar{x}_{1:k})^s \right]$$

$$= \sum_{\ell=1}^k \sum_{s=0}^p \binom{p}{s} \text{SP}_\ell^{p-s} (\bar{x}_\ell - \bar{x}_{1:k})^s$$

$$= \sum_{s=0}^p \binom{p}{s} \sum_{\ell=1}^k \text{SP}_\ell^{p-s} (\bar{x}_\ell - \bar{x}_{1:k})^s$$

$$= \sum_{s=0}^p \binom{p}{s} \sum_{\ell=1}^k \text{SP}_\ell^{p-s} \left( \bar{x}_\ell - \frac{1}{n} \sum_{i=1}^k n_i \bar{x}_i \right)^s,$$

which was to be shown. ∎

For convenience, we will split the proof of Theorem 5 into parts (a), (b) and (c) corresponding to the decompositions for the sums-of-squares, sums-of-cubes, and sums-of-quads. We will prove each of these in order, appealing to previously proved results as we go. Each result for the pooled sample can be obtained by application of Theorem 4, but we also show a useful alternative proof for part (a) using a decomposition of the centering matrix.

**PROOF OF THEOREM 5(A):** Applying Theorem 4 with $p = 2$ we obtain the preliminary step:

$$\text{SS}_{1:k} = \sum_{\ell=1}^k \text{SS}_\ell + \sum_{\ell=1}^k n_\ell (\bar{x}_\ell - \bar{x}_{1:k})^2 .$$

We then obtain the result in the theorem by observing that:

$$\text{SS}_{1:k} = \sum_{\ell=1}^k \text{SS}_\ell + \sum_{\ell=1}^k \frac{n_\ell}{n^2} (n\bar{x}_\ell - n\bar{x}_{1:k})^2$$

$$= \sum_{\ell=1}^k \text{SS}_\ell + \sum_{\ell=1}^k \frac{n_\ell}{n^2} \left( n\bar{x}_\ell - \sum_{i=1}^k n_i \bar{x}_i \right)^2$$

$$= \sum_{\ell=1}^{k} SS_\ell + \sum_{\ell=1}^{k} \frac{n_\ell}{n^2} \left[ n^2 \bar{x}_\ell^2 - 2n\bar{x}_\ell \left( \sum_{i=1}^{k} n_i \bar{x}_i \right) + \left( \sum_{i=1}^{k} n_i \bar{x}_i \right)^2 \right]$$

$$= \sum_{\ell=1}^{k} SS_\ell + \sum_{\ell=1}^{k} n_\ell \bar{x}_\ell^2 - \left( \frac{1}{n} \sum_{i=1}^{k} n_i \bar{x}_i \right) \sum_{\ell=1}^{k} n_\ell \left( 2\bar{x}_\ell - \frac{1}{n} \sum_{i=1}^{k} n_i \bar{x}_i \right)$$

$$= \sum_{\ell=1}^{k} SS_\ell + \sum_{\ell=1}^{k} n_\ell \bar{x}_\ell^2 - \left( \frac{1}{n} \sum_{i=1}^{k} n_i \bar{x}_i \right) \left( 2 \sum_{\ell=1}^{k} n_\ell \bar{x}_\ell - \frac{1}{n} \sum_{\ell=1}^{k} n_\ell \sum_{i=1}^{k} n_i \bar{x}_i \right)$$

$$= \sum_{\ell=1}^{k} SS_\ell + \sum_{\ell=1}^{k} n_\ell \bar{x}_\ell^2 - \left( \frac{1}{n} \sum_{i=1}^{k} n_i \bar{x}_i \right) \left( \sum_{\ell=1}^{k} n_\ell \bar{x}_\ell \right)$$

$$= \sum_{\ell=1}^{k} SS_\ell + \sum_{\ell=1}^{k} n_\ell \bar{x}_\ell^2 - \frac{1}{n} \left( \sum_{i=1}^{k} n_i \bar{x}_i \right)^2,$$

which was to be shown. (Note that one can also prove this result by induction using the formula for pooling of two subgroups in Theorem 4.) ∎

**ALTERNATIVE PROOF OF THEOREM 5(A):** Instead of using proof by induction, it is possible to prove the result as a consequence of a simple decomposition of the centering matrix. Suppose we let $n = n_1 + \cdots + n_k$ denote the pooled sample size and let $\mathbf{C}$ denote the $n \times n$ centering matrix. Using simple matrix manipulation we can get a decomposition for the centering matrix where we express this matrix in terms of the $n_i \times n_i$ centering matrices $\mathbf{C}_i$. A simple variation of a decomposition result shown in O'Neill (2020) (pp. 17-18) is:

$$\mathbf{C} = \begin{bmatrix} \mathbf{C}_1 & \mathbf{0} & \cdots & \mathbf{0} \\ \mathbf{0} & \mathbf{C}_2 & \cdots & \mathbf{0} \\ \vdots & \vdots & \ddots & \vdots \\ \mathbf{0} & \mathbf{0} & \cdots & \mathbf{C}_n \end{bmatrix} + \begin{bmatrix} \mathbf{1}_{n_1 \times n_1}/n_1 & \mathbf{0} & \cdots & \mathbf{0} \\ \mathbf{0} & \mathbf{1}_{n_2 \times n_2}/n_2 & \cdots & \mathbf{0} \\ \vdots & \vdots & \ddots & \vdots \\ \mathbf{0} & \mathbf{0} & \cdots & \mathbf{1}_{n_k \times n_k}/n_k \end{bmatrix} - \frac{\mathbf{1}_{n \times n}}{n}.$$

Now, let $\mathbf{x} = (\mathbf{x}_1, \mathbf{x}_2, \ldots, \mathbf{x}_k)$ denote the pooled data vector partitioned into its subgroups. The pooled sum-of-squares and the subgroup sums-of-squares can be written as:

$$SS_p = \mathbf{x}^T \mathbf{C} \mathbf{x} \qquad SS_i = \mathbf{x}_i^T \mathbf{C}_i \mathbf{x}_i.$$

Hence, we can obtain the sum-of-squares decomposition:

$$SS_{1:k} = \mathbf{x}^T \mathbf{C} \mathbf{x} = \sum_{i=1}^{k} \mathbf{x}_i^T \mathbf{C}_i \mathbf{x}_i + \sum_{i=1}^{k} \frac{\mathbf{x}_i^T \mathbf{1}_{n_i \times n_i} \mathbf{x}_i}{n_i} - \frac{\mathbf{x}^T \mathbf{1}_{n \times n} \mathbf{x}}{n}$$

$$= \sum_{i=1}^{k} \mathbf{x}_i^T \mathbf{C}_i \mathbf{x}_i + \sum_{i=1}^{k} \frac{(\mathbf{x}_i^T \mathbf{1}_{n_i})(\mathbf{1}_{n_i}^T \mathbf{x}_i)}{n_i} - \frac{(\mathbf{x}^T \mathbf{1}_n)(\mathbf{1}_n^T \mathbf{x})}{n}$$

$$= \mathrm{SS}_1 + \mathrm{SS}_2 + \cdots + \mathrm{SS}_k + \sum_{i=1}^{k} \frac{n_i^2 \bar{x}_i^2}{n_\ell} - \frac{(n\bar{x}_{1:k})^2}{n}$$

$$= \mathrm{SS}_1 + \mathrm{SS}_2 + \cdots + \mathrm{SS}_k + \sum_{i=1}^{k} n_i \bar{x}_i^2 - \frac{1}{n_1 + \cdots + n_k} \cdot \left(\sum_{i=1}^{k} n_i \bar{x}_i\right)^2,$$

which was to be shown. ∎

**PROOF OF THEOREM 5(B):** Applying Theorem 4 with $p = 3$ we obtain:

$$\mathrm{SC}_{1:k} = \sum_{\ell=1}^{k} \mathrm{SC}_\ell + 3 \sum_{\ell=1}^{k} \mathrm{SS}_\ell (\bar{x}_\ell - \bar{x}_{1:k}) + \sum_{\ell=1}^{k} n_\ell (\bar{x}_\ell - \bar{x}_{1:k})^3$$

$$= \mathrm{SC}_1 + \mathrm{SC}_2 + \cdots + \mathrm{SC}_k + 3 \sum_{\ell=1}^{k} \mathrm{SS}_\ell \left(\bar{x}_\ell - \frac{1}{n_1 + \cdots + n_k} \sum_{i=1}^{k} n_i \bar{x}_i\right)$$

$$+ \sum_{\ell=1}^{k} n_\ell \left(\bar{x}_\ell - \frac{1}{n_1 + \cdots + n_k} \sum_{i=1}^{k} n_i \bar{x}_i\right)^3,$$

which was to be shown. ∎

**PROOF OF THEOREM 4C:** Applying Theorem 4 with $p = 4$ we obtain:

$$\mathrm{SQ}_{1:k} = \sum_{\ell=1}^{k} \mathrm{SQ}_\ell + 4 \sum_{\ell=1}^{k} \mathrm{SC}_\ell (\bar{x}_\ell - \bar{x}_{1:k}) + 6 \sum_{\ell=1}^{k} \mathrm{SS}_\ell (\bar{x}_\ell - \bar{x}_{1:k})^2 + \sum_{\ell=1}^{k} n_\ell (\bar{x}_\ell - \bar{x}_{1:k})^4$$

$$= \mathrm{SQ}_1 + \mathrm{SQ}_2 + \cdots + \mathrm{SQ}_k + 4 \sum_{\ell=1}^{k} \mathrm{SC}_\ell \left(\bar{x}_\ell - \frac{1}{n_1 + \cdots + n_k} \sum_{i=1}^{k} n_i \bar{x}_i\right)$$

$$+ 6 \sum_{\ell=1}^{k} \mathrm{SS}_\ell \left(\bar{x}_\ell - \frac{1}{n_1 + \cdots + n_k} \sum_{i=1}^{k} n_i \bar{x}_i\right)^2$$

$$+ \sum_{\ell=1}^{k} n_\ell \left(\bar{x}_\ell - \frac{1}{n_1 + \cdots + n_k} \sum_{i=1}^{k} n_i \bar{x}_i\right)^4,$$

which was to be shown. ∎